# Understanding Social Networks using Transfer Learning

**Jun Sun**

*Institute for Web Science and Technologies (WeST), University of Koblenz–Landau, Koblenz, Germany, 56070*

**Steffen Staab**

*Institute for Web Science and Technologies (WeST), University of Koblenz–Landau, Koblenz, Germany, 56070*

*Web and Internet Science Research Group (WAIS), University of Southampton, UK, SO17 1BJ*

**Jérôme Kunegis**

*Namur Centre for Complex Systems (naXys), University of Namur, B-5000 Belgium*

*ABSTRACT. A detailed understanding of users contributes to the understanding of the Web's evolution, and to the development of Web applications. Although for new Web platforms such a study is especially important, it is often jeopardized by the lack of knowledge about novel phenomena due to the sparsity of data. Akin to human transfer of experiences from one domain to the next, transfer learning as a subfield of machine learning adapts knowledge acquired in one domain to a new domain. We systematically investigate how the concept of transfer learning may be applied to the study of users on newly created (emerging) Web platforms, and propose our transfer learning–based approach, TraNet. We show two use cases where TraNet is applied to tasks involving the identification of user trust and roles on different Web platforms. We compare the performance of TraNet with other approaches and find that our approach can best transfer knowledge on users across platforms in the given tasks.*

**EDITOR INTRO HEAD.**

EDITOR INTRO PARAGRAPH.

## 1. Introduction

The Web evolves in a permanent cycle (Fig. 1), as portrayed by Hendler et al. [1]: An idea may lead to novel technology as well as social activities. Taken together, the individual micro interactions of the many lead to meso and macro effects observable at a larger scale. Very often, new issues arise and people tinker with solutions starting the cycle from anew.

There has been and will be a multitude of Web platforms and many of them have repeated such a cycle of learning, often from scratch, sometimes based on anecdotal evidence contributed by experienced Web developers or community managers. In spite of long time learning from social scientists who studied behavior of individuals in groups, with abstracting lessons learned, little hard evidence could be brought to the table that might have been operationalized using quantification of user behavior.

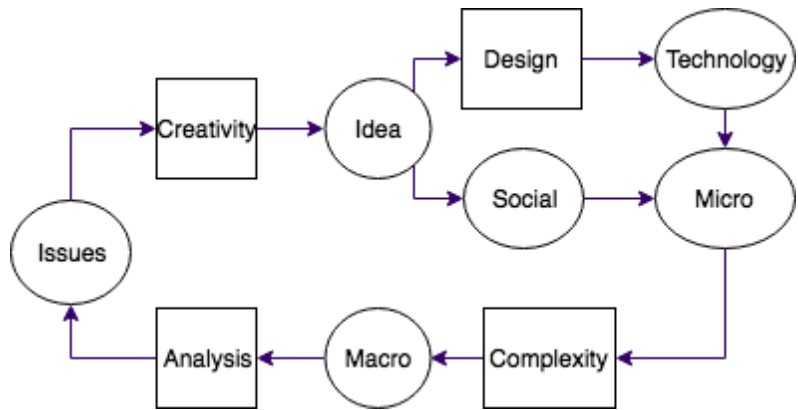

Figure 1. Web development cycle (adopted from Hendler et al. [1]).

Quantifiable cross-community analyses have been undertaken by, e.g., Rowe et al. [2]. They observed and classified multiple online communities. They proposed several measures useful to quantify differences between communities, such as the number of initiative-takers or the length of discussions. Considering these measures, however, it remained open whether and how they could actually be re-used to transfer experiences from one community to another. The evolution of a Web platform would be greatly facilitated, and the learning cycle would be cut short, if measurements of social behavior could be transferred from previous experiences to new ones, not just based on qualitative observations, but also based on quantifiable rules. For example, a new Web platform might want to discourage trolls and encourage trusted users without running through the learning cycle multiple times by transferring quantitative experiences from previous Web platforms.

The main challenge of learning from existing platforms lies in the fact that Web platforms are so heterogeneous in terms of size and structure. For example, a user with 20 friends in a small-scale network (e.g., a friendship network in a classroom) might be considered influential, while a user with the same number of friends in Facebook is far from being influential. Some platforms such as Slashdot contain negative user relationships such as "foes", and others do not.

Better than simply ignoring the heterogeneity, human experts are able to learn from few examples they observe from existing platforms, and transfer their "experience" to new situations. The analogy of such process in machine learning is called *transfer learning* [3], where we learn knowledge from *source* datasets we know well, and apply the knowledge to new *target* datasets.

In this paper, we specifically address the problem of transferring quantifiable measures of users across Web platforms. Such transfer is particularly challenging due to the above-mentioned heterogeneity of Web platforms. Thus, we extend previous studies and propose a transfer learning–based approach, TraNet, to benefit the evolution of Web platforms and reducing Web development efforts. TraNet is suitable when we have little knowledge on users of one platform, and wish to learn from existing, familiar platforms. In two case-studies, we show how TraNet can be applied to the study of users in different Web platforms. In the first example, we try to evaluate users' trustworthiness. In the second example, we try to identify users with specific roles.

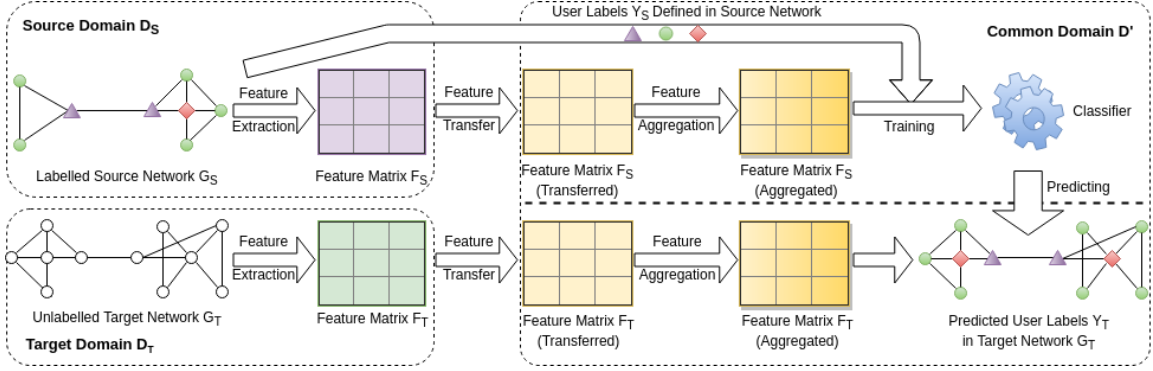

Figure 2. Overview of TraNet, the transfer learning procedure for user study described in this paper.

We briefly introduce the background knowledge on transfer learning and some related work in Section 2. In Section 3 we describe TraNet in detail and in Section 4 we evaluate two applications of TraNet. The necessary code (github.com/yfiua/TraNet) and public datasets to reproduce the results in the paper and apply TraNet to accomplish similar tasks are made available online.

## *2. Transfer Learning*

Conventionally, machine learning happens in one domain. For example, one may train a friend recommendation model with the data of one Web platform, and apply the trained model to recommend friends to users on the same platform. When we do not have enough data to train our model, "transferring" knowledge from other, existing platforms is necessary. However, the performance is not guaranteed if we directly use the trained model from other platforms. This is because the distribution of data from various platforms might differ considerably, and it violates the assumption in conventional machine learning that the training data from which we learn, and the future data to which we want to apply the knowledge we learnt are sampled from the same domain [3].

*Transfer learning* tackles this issue. Formally, we adopt the definition of transfer learning given by Pan and Yang [3]:

"Given a source domain $D_S$ and learning task $T_S$, a target domain $D_T$ and learning task $T_T$, transfer learning aims to help improve the learning of the target predictive function $f_T(\cdot)$ in $D_T$ using the knowledge in $D_S$ and $T_S$, where $D_S \neq D_T$ or $T_S \neq T_T$."

We use the term *source dataset* to refer to a dataset that belongs to the source domain $D_S$, that we learn knowledge from; and *target dataset* to refer to a dataset that belongs to the target domain $D_T$, to which we want to transfer the knowledge which we have learnt from the source dataset.

Researchers in the field of Web science have used transfer learning to tackle the data sparsity problem or the lack of ground truths (labels). For example, collaborative filtering aims to predict the interest of users, and can be used to give recommendation of products to customers in E-Commerce platforms for instance [4]. This problem can be seen as the prediction of missing values (to which extent a customer likes a potential item) in the adjacency matrix of a bipartite network. In order to mitigate the effect of data sparsity, Pan et al. [5] have proposed a transfer learning method based on *coordinate system transfer*, to effectively transfer knowledge about customers and products from other mature networks.

Transfer learning also helps infer social ties on the Web. Not all types of social ties are explicit on all platforms. For instance, negative links representing "foe" relationship are not present in Facebook or Twitter, but they might be inferred from platforms where negative links are present such as Slashdot-Zoo (see Section 4.1.1). Tang et al. [6] proposed a transfer learning–based algorithm (TranFG) to transfer high level social-psychological patterns from existing networks, such as the structural balance theory in signed networks and the status theory in networks where people have higher/lower statuses (e.g., advisors and advisees). Such learned patterns can then be used to infer social ties in other networks.

## *3. Proposed Method*

Fig. 2 illustrates our transfer learning approach, TraNet. TraNet comprises two phases: learning (top row) and inference (bottom row). The goal of the learning phase is to learn a model from the network of an existing Web platform with available user labels, and the goal of the inference phase is to apply the learned model to predict labels in other networks. In the example depicted in Fig. 2, we learn from the source dataset how to detect the following three predefined user labels: central users (diamonds), bridging users (triangles), and normal users (circles); and transfer the knowledge to the unlabeled target network to predict the user labels in it. The elements within the source domain $D_S$ and within the common domain $D'$ but above the dashed line are obtained during the learning phase, while the ones within the target domain $D_T$ and within the common domain $D'$ but below the dashed line are obtained during the inference phase.

With a network $G$ as input, we generate the feature matrix $F_{n\_m}$ (shown in Fig. 2 as grids), where $n$ is the number of nodes in $G$ and $m$ is the total number of features. Inside $F_{n\_m}$, each node (user) has $m$ structural features $x_j, (j \in \{1, 2, ..., m\})$, which are expected to be non-domain-specific. The feature generation process consists of three steps: (i) feature extraction, (ii) feature transfer by feature transformation, and (iii) feature aggregation. They will be explained in the following subsections.

In the source dataset, users as nodes in the network are labelled with a vector $Y_S$ of length $n_S$, where $n_S$ is the number of users in the source network. Each value $y_i \in \{1, 2, ..., k\}, (i \in \{1, 2, ..., n_S\})$ in $Y_S$ denotes the label of the *i*-th user, where $k$ is the total number of possible labels. In the target dataset, the label vector $Y_T$ is to be predicted.

With $Y_S$ and the feature matrices $F_S$, $F_T$ for both source and target networks, the label prediction problem reduces to a classification problem. In the learning phase, we optimize a predictive function $y_{predict} = f(x_1, x_2, ... , x_m)$ that maps a node's features $[x_1, x_2, ... , x_m]$ into a label $y_{predict} \in \{1, 2, ..., k\}$ which matches the corresponding value in $Y_S$.

Such a predictive function *f* can be regarded as a user classifier which predicts user labels in the common domain, since the features in $F_S$ are expected to be non-domain-specific, and are already transformed in a way that they can match across networks. In the inference phase, we use $f$ to compute $y_{predict}$ for all nodes in $G_T$ in order to predict user labels in the target network.

In our implementation, we use a random forest classifier. Once the classifier is trained, it can be saved and predict the probability that each node (user) belongs to each class (label) in future datasets, given the network structure as input.

In the inference phase, once we obtain the feature matrix $F_T$ for the target dataset, we can use the pre-trained classifier to predict the user labels in the target network.

The rest of the section describes the individual steps for feature generation.

## 3.1 Feature Extraction

Users behave in different ways on the Web. A user's behavior is reflected in her surrounding network structure, and thus we can examine the structural features of a node to study the user. For instance, we can group users with similar behavior into a *role* and examine the common pattern appearing in their neighborhood. Users with a similar neighborhood pattern can then be classified into the same role [7].

Structural features of nodes can be extracted by only looking at the structure (e.g., the adjacency matrix) of the network, without requiring information on additional attributes of nodes or links (e.g., users' geolocations as node attributes, or message contents as link attributes in a user interaction network) [8]. In machine learning, it would also be useful to take these additional attributes into consideration in order to improve the performance. However, these node and link attributes are usually domain-specific, and may not be applicable in other networks. In transfer learning, blindly transferring knowledge may not be successful, or even make the performance of learning worse [3]. Since structural features are by nature present in all networks, in order to better investigate the principle idea behind transfer learning on the Web, we focus on the aspect of structural features.

### 3.1.1 Base Feature Extraction

For each node in a given network, we compute the following five structural features as its *base features*: (i) degree (ii) indegree (iii) outdegree (iv) local clustering coefficient (v) PageRank.

### 3.1.2 Feature Aggregation

To characterize a user on a Web platform, it is important to know not only by who she is, but also who she knows and who knows her. In terms of machine learning, we do not only consider a node's local features, but also look into its neighborhood's features and the network structure around it. Inspired by the idea of *recursive features* proposed in [8], for each node in the network, we generate its *neighborhood features* by aggregating its neighbors' features step by step. For more details, in the first round, for each node and each local feature, we compute the average feature value of its neighbors and store it as a new feature. In the following rounds, we aggregate the features that we get in the last round in the same way.

As to the total number of rounds $r_{max}$ for which we perform the above described repetitive feature aggregation process, we choose $r_{max} = 5$ in practice. Considering the de facto low diameters of real-world networks, this provides us a good trade-off between classification performance and computational overhead.

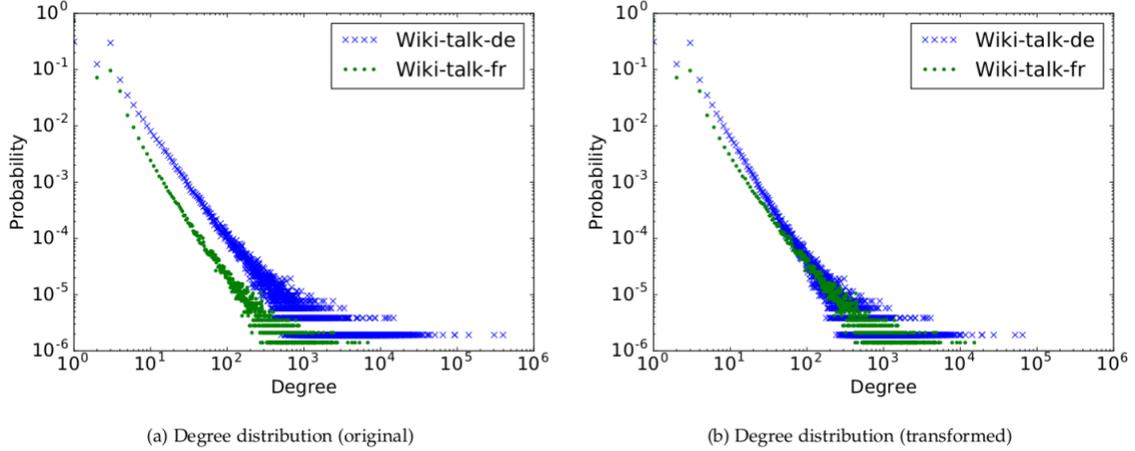

(a) Degree distribution (original)   (b) Degree distribution (transformed)

Figure 3. Degree distributions of the German (`de`) and French (`fr`) `Wiki-Talk` networks (see Section 4.1.1) before and after power-law degree transformation. Each dot in the plot represents the probability (Y axis) of a degree value (X axis) in the network. Two separated curves overlap after the transformation.

### 3.2 Feature Transfer by Feature Transformation

The main challenge in transfer learning is that the distributions of features differ between the source and target datasets. Thus, features that are extracted from different networks are often not directly comparable. Therefore, after all base features are extracted, we transform them via different methods in order to make them comparable across networks. The transformation of features to a dataset-independent space of values is performed separately for each dataset.

Feature transformation is especially difficult since the target dataset might not be seen during the training phase [9]. In our approach, the general idea of feature transformation is to define a common feature distribution for each kind of base feature, which is more likely to be comparable across networks. Therefore, the feature transformation procedure is network-independent and order-free, i.e. we do not need to access the target network when we perform feature transformation for the source network, and vice versa.

We discuss the following transformation methods for our base features.

### 3.2.1 Power-law Degree Transformation

Studies have shown that some base features such as nodes' degree approximately follow power-law distribution in real networks [10]:

$$p(x) = c \cdot x^{-\alpha} \qquad \alpha > 1, x \in [x_{min}, +\infty), x_{min} > 0$$

*(Eq. 1)*

Given this prior knowledge, we can transform the feature values according to their *quantile* values (Equation 2). The quantile of a feature value $x$ is defined as the probability that any value in its domain is less than $x$, i.e., $\int_{x_{min}}^{x} p(x)dx$.

$$\int_{x_{min}}^{x} p(x)dx = \int_{x'_{min}}^{x'} p'(x')dx'$$

*(Eq. 2)*

By these means, we can transform any kind of power-law like feature distribution into one

common power-law distribution. We choose the power-law distribution:
$$p'(x') = x'^{-2} \qquad x' \in [1, +\infty)$$
(Eq. 3)

as the target distribution of transformation for the ease of calculation. Combining Equations 1, 2 and 3, we get:
$$x' = (\frac{x}{x_{min}})^{\alpha-1}$$
(Eq. 4)

where $x'$ is the transformed feature value. Considering our scenario where degree $d$ commonly starts from 1, the formula can be further simplified to:
$$x' = d^{\alpha-1}$$
(Eq. 5)

In our implementation, we use the method by Clauset et al. [10] to fit a power-law distribution and estimate the exponential $\alpha$.

Fig. 3 shows the degree distributions of two networks before and after our transformation. The original curves of degree distributions are clearly separated (as in Fig. 3a), while being overlapping after the transformation (as in Fig. 3b). This indicates that our transformation method can approximately transform degree distributions from different networks into a common power-law distribution.

### 3.2.2 PageRank Transformation

The standard PageRank of nodes in a network is defined as the stationary probability distribution in a converged random surfing process, and is often used to measure the centrality of nodes. However, it can be biased by the network size. Berberich et al. have proposed the *normalized PageRank* [11], where standard PageRank values are normalized by their theoretical lower bound. The normalized PageRank has been proved to be independent of network size and comparable across networks. Hence, we use it as the transformation for our base feature PageRank.

## *4. Applications*

We now illustrate two concrete applications of TraNet as examples: identifying trusted users in social networks on the Web (in Section 4.2) and identifying users with specific roles (in Section 4.3). To evaluate the performance of TraNet, we accomplish the same tasks with other approaches and compare the performance.

### 4.1 Settings

We now present our application settings.

#### 4.1.1 Datasets

We apply TraNet to the following real-world datasets from the Web.
- `Slashdot-Zoo` is a signed network dataset extracted from Slashdot, consisting of 79,120 users and 515,397 directed relations [12]. In this network, each directed signed edge represents a "friend" (positive) or "foe" (negative) relation from one user to another on the technology news site Slashdot, where each user can explicitly mark other users as their friends or foes in order to increase or decrease the chance to see their posts.
- `Epinion-Trust` is a signed network of Epinions, an online product rating site

[13]. It consists of 131,828 users and 841,372 directed, signed edges, each representing a trust (positive) or distrust (negative) relation from one user to any user (possibly herself).
- `ARIS` contains the user interaction network in the ARIS Community, the internal Business Process Management (BPM) system used in Software AG company, the second largest software company in Germany. At the time we extracted it, it had 9,566 threads and 20,538 comments by 4,216 users, among which 885 were labelled as trusted users. We use a directed edge to represent a user's comment to another user's post or comment.
- `Wiki-Talk` is a set of user interaction networks in Wikipedia of different languages. In Wikipedia, users can communicate with each other on their talk pages. We extract the interactions on all users' talk pages of Wikipedia in different languages (each language forms an individual network). We use one node to represent a registered Wikipedia user, and one directed edge to represent a user interaction. Additionally, some users are labelled as administrators (`Admin`) by the Wikipedia community, among the others `Normal` users. A more detailed description of `Wiki-Talk` can be found online [14, 15].

#### 4.1.2 Baselines
We choose the following baselines to compare the performance of our approach (denoted as `TraNet`). We have also tried approaches such as the transfer component analysis (TCA) [16], but have found that they do not scale to suit our applications.
- `None`: training a model from the source network and directly applying it to the target network. This serves as a lower-bound baseline, since no feature transformation is done.
- `Trad.`: traditional machine learning (i.e. training a model from partial data in a network and apply it to the rest data in the same network). This serves as an upper-bound baseline, since training and test data are sampled from the same domain, and no transformation is necessary.
- `SVD`: performing feature transformation based on the singular value decomposition (SVD) proposed by Agirre and De Lacalle [17].
- `TrAda.`: performing transfer learning with TrAdaBoost proposed by Dai et al. [18]. TrAdaBoost has a different setting from ours: it requires partially labelled data from the target network

#### 4.1.3 Implementation
We implement `TraNet` with Python. We use our own implementation for `SVD`, and adopt an open source implementation (`github.com/chenchiwei/tradaboost`) for `TrAda`. We use the ROC-AUC metric to measure the performance of the classifiers.

### 4.2 Application in Trust Transfer
Now we apply TraNet to predict trusted users on Web platforms. We compute the trusted users in `Slashdot-Zoo` and `Epinion-Trust` using the EigenTrust algorithm [19], and use each of them as the source dataset to learn a model respectively, and predict the trusted users in `ARIS`. The result is shown in Table 1. It shows that performing no feature

### Table 1
*Predicting trusted users in the target network ARIS with the knowledge transferred from the two source networks Slashdot-Zoo and Epinion-Trust respectively. The values in the table show the ROC-AUC performance of the classifier in different settings (see Section 4.1.2). We can achieve the best performance with transfer learning using our approach TraNet.*

|  | Source Dataset | |
| --- | --- | --- |
|  | Slashdot-Zoo | Epinion-Trust |
| None | 0.7629 | 0.6755 |
| SVD | 0.6090 | 0.6864 |
| TrAda. | 0.6698 | 0.6889 |
| **TraNet** | 0.8255 | 0.7500 |
| Trad. | 0.8592 | |

transformation does not work well. Our approach TraNet outperforms other transfer learning approaches, and can even achieve the performance close to traditional within-network learning when using Slashdot-Zoo as the source dataset.

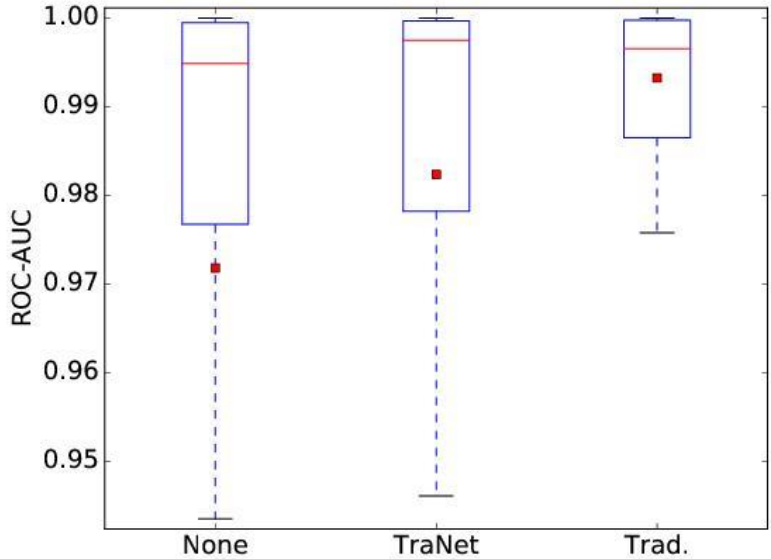

Figure 4. ROC-AUC performance of the Admin classifier in different settings (see Section 4.1.2). In each box plot, the red bar shows the median value, while the red dot shows the mean value of the ROC-AUC in each experiment. SVD and TrAda. are omitted here due to their poor performance (0.782 and 0.605 on average, respectively). Our approach TraNet achieves an average ROC-AUC of 0.982, which is the best among all transfer learning approaches.

### 4.3 Application in Role Transfer

In this application, we use the Wiki-Talk datasets and try to identify administrators among normal users.

For each of the 14 sub-datasets in `Wiki-Talk` which contains at least 25 `Admins`, we build one binary classifier for `Admins`. Each of the classifiers is applied to the other 13 sub-datasets. Therefore, we have 182 pairs of source and target datasets.

As shown in Fig. 4, high ROC-AUC (0.997 on average) with traditional machine learning (`Trad.`) indicates that identifying Admins is achievable with given data. Transfer learning with `SVD` and `TrAda.` is ineffective with decreased performance compared with None. Our transfer learning approach can achieve the best performance with an average ROC-AUC of 0.982, improving by more than 1% compared with the transfer learning without feature transformation (`None`).

## *5. Conclusion*

We have proposed a transfer learning–based approach, TraNet, to study users on the Web. It provides a novel method to transfer measurements of users' social behavior (i.e. labels) from existing platforms in order to better analyze social effects on, especially, new Web platforms, and thus helps cut short the learning cycle of Web development as shown in the very beginning of the paper.

We have been focusing on the study of users. TraNet can potentially be applied to study other entities on the Web such as groups or products, since they can also be represented as nodes in the network. This shows a direction of future work.

**Acknowledgements**
*The authors would like to thank Software AG for providing the dataset. The research leading to these results has received funding from the European Community's Horizon 2020 - Research and Innovation Framework Programme under grant agreement No. 770469, CUTLER.*

**Jun Sun** is scientific researcher at Univ. of Koblenz-Landau. He graduated at TU Dresden in the field of computer science in 2015. He is interested in machine learning techniques applied in Web and network science.

**Steffen Staab** heads the Institute for Web Science and Technologies at Univ. of Koblenz– Landau and holds a chair for Web and Computer Science at Univ. of Southampton. He is interested in the many ways in which semantics of communication and data emerges and is used in nature and in IT.

**Jérôme Kunegis** is postdoctoral researcher at the Namur Center for Complex Systems (naXys) at Univ. of Namur, Belgium. Dr. Kunegis graduated at TU Berlin in the field of computer science in 2006, and received his PhD in 2011 at Univ. of Koblenz–Landau, Germany, on the spectral analysis of evolving networks.